\documentclass{article}

\usepackage{arxiv}
\usepackage[utf8]{inputenc} 
\usepackage[T1]{fontenc}    
\usepackage{hyperref}       
\usepackage{url}            
\usepackage{booktabs}       
\usepackage{amsfonts}       
\usepackage{nicefrac}           
\usepackage{lipsum}
\usepackage{graphicx}
\usepackage{colortbl}
\usepackage{xcolor}
\usepackage{array}
\usepackage{amsmath}
\usepackage{setspace}
\usepackage{appendix}
\graphicspath{ {./images/} }
\title{Predicting VCSEL Emission Properties Using Transformer Neural Networks}

\author{
  Aleksei V. Belonovskii$^{1,*}$, Elizaveta I. Girshova$^1$, Erkki Lähderanta$^1$, Mikhail Kaliteevski \\
  \\
  $^1$Lappeenranta University of Technology LUT, Skinnarilankatu 34, 53850, Lappeenranta, Finland \\
  \\
  $^*$Corresponding author: \texttt{aleksei.belonovskii@gmail.com}
}

\begin{document}
\maketitle
\begin{abstract}
This study presents an innovative approach to predicting VCSEL emission characteristics using transformer neural networks. We demonstrate how to modify the transformer neural network for applications in physics. Our model achieved high accuracy in predicting parameters such as VCSEL's eigenenergy, quality factor, and threshold material gain, based on the laser's structure. This model trains faster and predicts more accurately compared to traditional neural networks. The transformer architecture we propose is also suitable for applications in other fields. A demo version is available for testing at \url{https://abelonovskii.github.io/opto-transformer}.
\end{abstract}


\section{Introduction}
The concept of vertical-cavity surface-emitting lasers (VCSELs) was pioneered by Kenichi Iga who, along with his team at the Tokyo Institute of Technology, proposed and demonstrated the fundamental design in the early 1980s \cite{Iga1998}. The main advantage of VCSELs over traditional edge-emitting lasers is their ability to emit light perpendicular to the surface of the semiconductor wafer. This characteristic enables more compact and efficient optical devices. VCSELs are widely used in applications such as optical communications, sensing, and high-speed data transfer, noted for their high efficiency and the ability to be fabricated in arrays \cite{Padullaparthi2021}.

VCSELs consist of several thin layers, often using a GaAs-AlGaAs system for their lattice-matched properties which allow efficient layering on a GaAs substrate. This system also varies in refractive index with changes in aluminum content, enhancing the Bragg mirror's efficiency and allowing high aluminum concentrations to form an oxide that aids in current confinement, enabling low threshold currents. For growth techniques, advancements include a commercial process for oxide VCSELs that focuses on uniform epitaxial thickness and aperture control, leading to high-yield, low-voltage operation with promising reliability comparable to proton-implanted VCSELs \cite{Lei1999}. Significant research includes the development of a conductive distributed Bragg reflector (DBR) using a single material indium tin oxide demonstrating notable index contrast and reflectivity improvements through porous layers \cite{Schubert2007}. Moreover, VCSELs are being integrated onto biocompatible PDMS substrates, showing potential in biomedical fields due to their effective low threshold and high output \cite{Kwon2023}. Modern VCSELs feature low threshold current densities and high output powers, making them excellent for short-range optical communications and optical interconnects \cite{Huffaker1999,Li2021,Eitel2000}.

In calculating VCSEL parameters, various methods are applied to ascertain optical and electrical characteristics, including electric field distribution inside the resonator, reflection and transmission coefficients, eigenmodes, threshold values, etc. Analytically, the Transfer-Matrix Method (TMM) is utilized to calculate reflection and transmission coefficients of multilayer structures and to find eigenenergies \cite{Born1999}. Numerically, methods such as Finite-Difference Time-Domain (FDTD), Finite Element Method (FEM), and Eigenmode Expansion (EME) provide detailed insights into the complex geometries of nanostructures \cite{Wilmsen2001}.

Traditional VCSEL modeling methods often involve complex simulations and empirical models, requiring extensive computational resources and in-depth knowledge of the underlying physical processes. These methods can be time-consuming and computationally intensive, as each structural modification necessitates a new simulation, further increasing the time and complexity involved. Therefore, developing faster modeling techniques, such as neural networks, is crucial for efficient VCSEL design and analysis.

The recent advancements in neural networks and machine learning architectures have had a significant impact on various fields, including natural language processing, computer vision, and, more recently, the modeling of complex physical systems. Artificial neural networks (ANNs) were used to approximate the electromagnetic responses of complex plasmonic waveguide-coupled with cavities structure (PWCCS) \cite{Zhang2019}. The ANNs used in this study were optimized by varying hyperparameters such as the number of layers, neurons per layer, solvers for weight optimization, and activation functions for the hidden layers by genetics algorithms (GA). The study demonstrates that ANNs can significantly improve the efficiency of spectrum prediction and inverse design for PWCCSs. The results indicate that the trained ANNs can predict the electromagnetic spectrum with high precision using only a small sampling of simulation results. The use of ANNs, optimized via GA, provides a more efficient method for both spectrum prediction and inverse design in PWCCSs, addressing key challenges in computational cost and data consistency, thus advancing the design and optimization of complex photonic devices.

There are several examples of feedforward neural networks for photonics \cite{Han2021, Ma2020}. They were used to learn electromagnetic scattering with variable thicknesses and materials, proposing a tandem architecture for forward simulation and inverse design \cite{Liu2018}. Peurifoy et al. used a feedforward neural network with four hidden layers to approximate light scattering in core-shell nanoparticles. Malkiel et al. developed a neural network with multiple layers for direct on-demand engineering of plasmonic structures \cite{Malkiel2018}. Tahersima et al. built a robust deeper network based on feedforward neural network for inverse design of integrated photonic devices, utilizing intensity shortcuts from deep residual networks \cite{Tahersima2019}. Generative adversarial networks (GAN) are usually used for generating patterns, particularly in tasks involving design optimization of metasurfaces and other photonic structures. Liu et al. combined a GAN with a simulation network to optimize metasurface patterns \cite{Liu2018GAN}. Jiang et al. used GANs to produce high-efficiency, topologically complex devices for metagrating design \cite{Jiang2019}.

In \cite{Asano2018} the authors present a method to optimize the Q factors of two-dimensional photonic crystal (2D-PC) nanocavities using deep learning. The authors created a dataset of 1000 nanocavities with randomly displaced air holes and calculated their Q factors using first-principles methods. They trained a four-layer neural network, including a convolutional layer, to understand the relationship between air hole displacements and Q factors. The neural network could predict Q factors with a standard deviation error of 13\% and estimate the gradient of Q factors quickly using back-propagation. This allowed them to optimize a nanocavity structure to achieve an extremely high Q factor of $1.58 \times 10^9$ over $\sim10^6$ iterations, significantly surpassing previous optimization methods. It was shown that the deep learning-based approach provides a highly efficient method for optimizing the Q factors of 2D-PC nanocavities, making it feasible to explore large parameter spaces and achieve significantly higher Q factors than previously possible with traditional optimization methods.

These insights showcase the potential of integrating machine learning techniques with photonic system design to enhance performance and efficiency. One such breakthrough is the Transformer architecture, which has demonstrated exceptional capabilities in handling sequential data. Introduced in 2017 by Vaswani et al. \cite{Vaswani2017}, the Transformer utilizes an "attention mechanism" that allows the neural network to focus on specific details and assess some parameters in relation to others. Deep learning has already proven its efficacy in analyzing, interpolating, and optimizing complex phenomena in various fields, including robotics, image classification, and language translation. However, the application of neural networks in photonics is still in its early stages, and there are many challenges that this approach encounters. 

Recently, researchers have started to use deep learning to accelerate the simulation process, leveraging its strong generalization ability, including Transformer \cite{Ma2023, Soroush2023, Yan2022}. Although obtaining the training dataset through physical simulation can take some time, such an investment is a one-time expenditure. Once trained, these neural networks are capable of capturing the general mapping between the space of structures and the space of optical properties, serving as a fast and computationally efficient surrogate model to replace physical simulations.

Our work is aimed at adapting the transformer architecture for tasks in mathematics and physics. We used embeddings that indicate the physical properties and conditions of the object, shifting the transformer into the physical plane. As an application demonstrating our transformer's capabilities, we chose the task of predicting VCSEL emission parameters. Our approach combines physical and trainable embeding components, uses positional encoding to preserve order information, and a multi-layer decoder structure to process input data and generate accurate predictions. The flexibility and adaptability of the transformer allow our model to successfully work with various types of laser structures and operating conditions, applying the architecture to model VCSEL, including its components such as DBR and a single layer, which we call here as Fabry-Pérot. We aim to provide a reliable and flexible tool that can accurately predict key emission parameters, facilitating the design and optimization of advanced photonic systems. Our work opens up new possibilities for using advanced neural network architectures in the field of photonics and beyond.

\section{Models of Nanostructures}
\label{sec:models_of_nanostructures}

The development of our training dataset was guided by the methodologies outlined in reference \cite{Bienstman2001}, which provided a standard VCSEL model framework. We generated approximately 150,000 distinct models across three primary categories: Fabry-Pérot structures (one layer in the medium), distributed Bragg reflectors (DBRs), and complete VCSEL configurations, utilizing the COMSOL Multiphysics software for simulation.

\begin{figure}[htbp]
    \centering
    \includegraphics[width=1\textwidth]{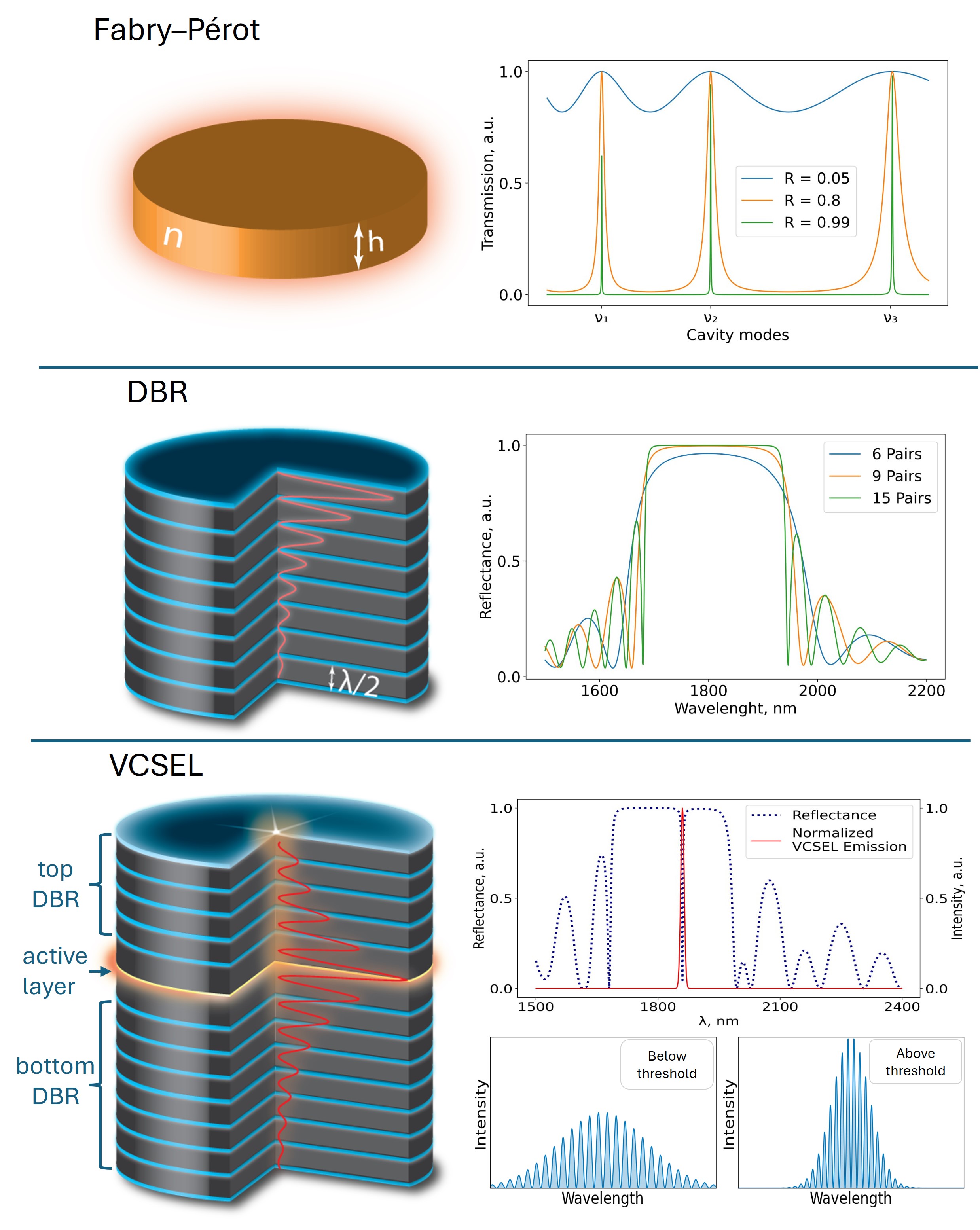}
    \caption{The figure illustrates key components and typical spectra associated with various nanostructures: \textbf{Top}: Fabry-Pérot - shows the active layer and a typical transmission spectrum for different mirror reflectivities. \textbf{Middle}: DBR - depicts the DBR structure and its reflectance spectrum for varying layer pairs. \textbf{Bottom}: VCSEL - displays the VCSEL structure and typical reflectance and emission spectra below and above the lasing threshold.}
    \label{fig:nanostructures}
\end{figure}

Figure \ref{fig:nanostructures} provides a comprehensive visual representation of several nanostructures and their typical spectral responses:

\begin{enumerate}
    \item \textbf{Top Section: Fabry-Pérot}
    \begin{itemize}
        \item In the top portion of the figure, the Fabry-Pérot is depicted. It consists of an active layer, which is responsible for the light interaction.
        \item To the right of the diagram, there is a typical transmission spectrum of the Fabry-Pérot interferometer. This spectrum illustrates how the transmission characteristics change with varying reflectivity levels of the mirrors. Peaks in the spectrum correspond to wavelengths that resonate within the cavity, while troughs represent non-resonant wavelengths.
    \end{itemize}
    
    \item \textbf{Middle Section: Distributed Bragg Reflector (DBR)}
    \begin{itemize}
        \item The middle section of the figure features a Distributed Bragg Reflector (DBR). This structure comprises alternating layers of materials with different refractive indices. These layers are designed to reflect specific wavelengths of light through constructive interference.
        \item To the right, the reflectance spectrum of a typical DBR is shown. This spectrum indicates how the reflectance varies with the number of layer pairs in the DBR. More pairs result in higher reflectivity, which is crucial for creating highly reflective mirrors in optical devices.
    \end{itemize}
    
    \item \textbf{Bottom Section: Vertical-Cavity Surface-Emitting Laser (VCSEL)}
    \begin{itemize}
        \item In the bottom section, the figure presents a Vertical-Cavity Surface-Emitting Laser (VCSEL). The VCSEL is constructed with a gain medium positioned between two DBRs, forming a vertical optical cavity. This design allows the laser to emit light perpendicular to the surface of the device.
        \item To the right of the VCSEL diagram, two spectra are displayed. The first spectrum is a typical reflectance spectrum, showing how the DBRs reflect light at specific wavelengths. The second spectrum illustrates the emission characteristics of the VCSEL, showing how the device emits light both below and above the lasing threshold.
    \end{itemize}
\end{enumerate}

To enhance our understanding of VCSELs' physical dynamics, we employed 2D axial models within COMSOL to precisely compute eigenmodes and other essential characteristics. These models spanned a broad spectrum of VCSEL designs by varying parameter values, including the real and imaginary components of the refractive index, and layer thicknesses. Materials commonly used in VCSEL construction such as \textit{GaAs}, \textit{AlAs}, \textit{AlGaAs}, \textit{TiO}$_2$, \textit{SiO}$_2$, \textit{Si}, and \textit{Si}$_3$\textit{N}$_4$ were incorporated into our simulations.

For the single-layer models, a range of initial energy values was used for the calculations. The real refractive index ($n$) and imaginary refractive index ($k$) values, as well as the thickness of each layer, were varied extensively. For the DBR and VCSEL structures, the $n$ and $k$ values were chosen to match the material pairs used and were dependent on the specified initial energy ranges for the calculations. The layer thicknesses were adjusted to ensure that each layer had an optical thickness equal to a quarter-wavelength or a half-wavelength, optimizing the optical properties. Moreover, initial energy values for DBR and VCSEL did not exceed two electron volts to avoid increased material dissipation, which is common in most materials at higher energy levels.

Furthermore, the number of layer pairs in the DBRs varied extensively, from one to fifty pairs in pure DBR models and between twenty to forty-five pairs in the upper and lower DBRs of VCSEL models. In VCSELs, the number of pairs in upper layers was always less than in the lower layers to optimize the structural symmetry and performance. The thickness of the quantum wells was also varied between $5$ to $15$ nm to study different energy configurations, primarily focusing on models where the threshold material gain was not excessively high to better understand effective laser models. For the data types involving only DBRs, no active layers were added, thus no threshold material gain calculations were needed for these types.

To accurately model and predict the physical parameters of these nanostructures, we structured the input embeddings to capture the main physical properties of the nanostructures. These embeddings specify key quantities such as layer thickness, refractive index (both real and imaginary components), and specific conditions like boundary or DBR layers. Each physical embedding is a vector containing detailed information about these physical properties, allowing the model to retain critical quantitative details that are essential for accurate simulations.

The table \ref{tab:physical_embeddings} shows an example of how these physical embeddings are structured.

\begin{table}[ht]
    \centering
    \caption{Physical embeddings for DBR.}
    \label{tab:physical_embeddings}
    \renewcommand{\arraystretch}{1.2} 
    \setlength{\arrayrulewidth}{0.3mm} 
    \setlength{\tabcolsep}{4pt}
    \newcolumntype{C}[1]{>{\centering\arraybackslash}m{#1}} 
    
    \begin{tabular}{C{0.5cm} | C{1cm} C{1.5cm} C{1cm} C{1cm} C{1cm} C{1.5cm} C{1cm} C{1cm} C{2cm}}
    
        & \textit{Energy} & \textit{Thickness} & \textit{N} & \textit{K} & \textit{GAIN} & \textit{Boundary} & \textit{DBR} & \textit{Pairs} & \textit{Value} \\ \hline
        \rowcolor{green!10} 1 & 1 & 0 & 0 & 0 & 0 & 0 & 0 & 0 & 1.2 \\
        2 & 0 & 0 & 1 & 0 & 0 & 1 & 0 & 0 & 1.0 \\
        \rowcolor{green!10} 3 & 0 & 0 & 0 & 1 & 0 & 1 & 0 & 0 & 0.0 \\
        4 & 0 & 0 & 0 & 0 & 0 & 0 & 1 & 1 & 40 \\
        \rowcolor{green!10} 5 & 0 & 1 & 0 & 0 & 0 & 0 & 1 & 0 & 136.49 \\
        6 & 0 & 0 & 1 & 0 & 0 & 0 & 1 & 0 & 3.46 \\
        \rowcolor{green!10} 7 & 0 & 0 & 0 & 1 & 0 & 0 & 1 & 0 & 0.0 \\
        8 & 0 & 1 & 0 & 0 & 0 & 0 & 1 & 0 & 15.03 \\
        \rowcolor{green!10} 9 & 0 & 0 & 1 & 0 & 0 & 0 & 1 & 0 & 2.95 \\
        10 & 0 & 0 & 0 & 1 & 0 & 0 & 1 & 0 & 0.0 \\
        \rowcolor{green!10} 11 & 0 & 0 & 1 & 0 & 0 & 1 & 0 & 0 & 1.0 \\
        12 & 0 & 0 & 0 & 1 & 0 & 1 & 0 & 0 & 0.0 \\ 
    \end{tabular}
\end{table}

The table \ref{tab:physical_embeddings} shows 12 embeddings for the Distributed Bragg Reflector (DBR). These embeddings are designed to capture key physical properties relevant to the DBR layers used in VCSELs:

\begin{enumerate}
    \item \textbf{First Five Columns (Energy, Thickness, N, K, GAIN)}:
    These columns determine the nature of the physical quantity.
    \begin{itemize}
        \item \textbf{Energy}: Indicates whether the row represents energy parameters.
        \item \textbf{Thickness}: Specifies the layer thickness in nanometers.
        \item \textbf{N}: Represents the real part of the refractive index.
        \item \textbf{K}: Denotes the imaginary part of the refractive index.
        \item \textbf{GAIN}: For active layers, the gain is represented instead of K.
    \end{itemize}
    
    \item \textbf{Boundary}:
    \begin{itemize}
        \item This column indicates whether the parameter pertains to the upper or lower part of the structure.
    \end{itemize}
    
    \item \textbf{DBR Indicator}:
    \begin{itemize}
        \item This column specifies whether the value is related to the DBR layer pairs.
    \end{itemize}
    
    \item \textbf{Pairs}:
    \begin{itemize}
        \item Indicates the number of layer pairs in the DBR.
    \end{itemize}
    
    \item \textbf{Value}:
    \begin{itemize}
        \item Represents the specific value associated with the physical property described by the embeddings in the row.
    \end{itemize}
\end{enumerate}

For instance, a row with a '1' in the "Thickness" column and '0's in the other first five columns indicates that the embedding represents a thickness value. The "Boundary" column would then show if this thickness pertains to the upper or lower layer, and the "DBR" columns indicate if this thickness is associated with a DBR layers.

This type of embedding schema allows for clear and convenient identification of physical quantities across all types of nanostructures under consideration. Each embedding vector uniquely encodes specific physical parameters, making it easier to systematically represent and analyze diverse attributes such as energy, thickness, refractive indices, and layer properties.

For example, from Table \ref{tab:physical_embeddings}, we can precisely deduce that we are examining the parameters of the DBR for energy at $1.2$ $eV$ on the lower boundary with $n=1$ and $k=0$. It specifies $40$ layer pairs. The first layer in the pair has a thickness of $136.49$ nm, $n=3.46$, and $k=0$. The second layer in the pair has a thickness of $15.03$ nm, $n=2.95$, and $k=0$. The parameters for the upper semi-infinite medium are $n=1$ and $k=0$.

By including these embeddings, the model is equipped to handle the complex relationships between physical properties and the resulting optical behaviors of nanostructures. The detailed configurations for other cases are provided in Appendix \ref{app:other_embeddings}.

This detailed capture of physical parameters through embeddings is a crucial step before delving into the transformer model's architecture, described in the following section.

\section{Transformer Model Description}
\label{sec:transformer_model}

\begin{figure}[htbp]
    \centering
    \includegraphics[width=\textwidth]{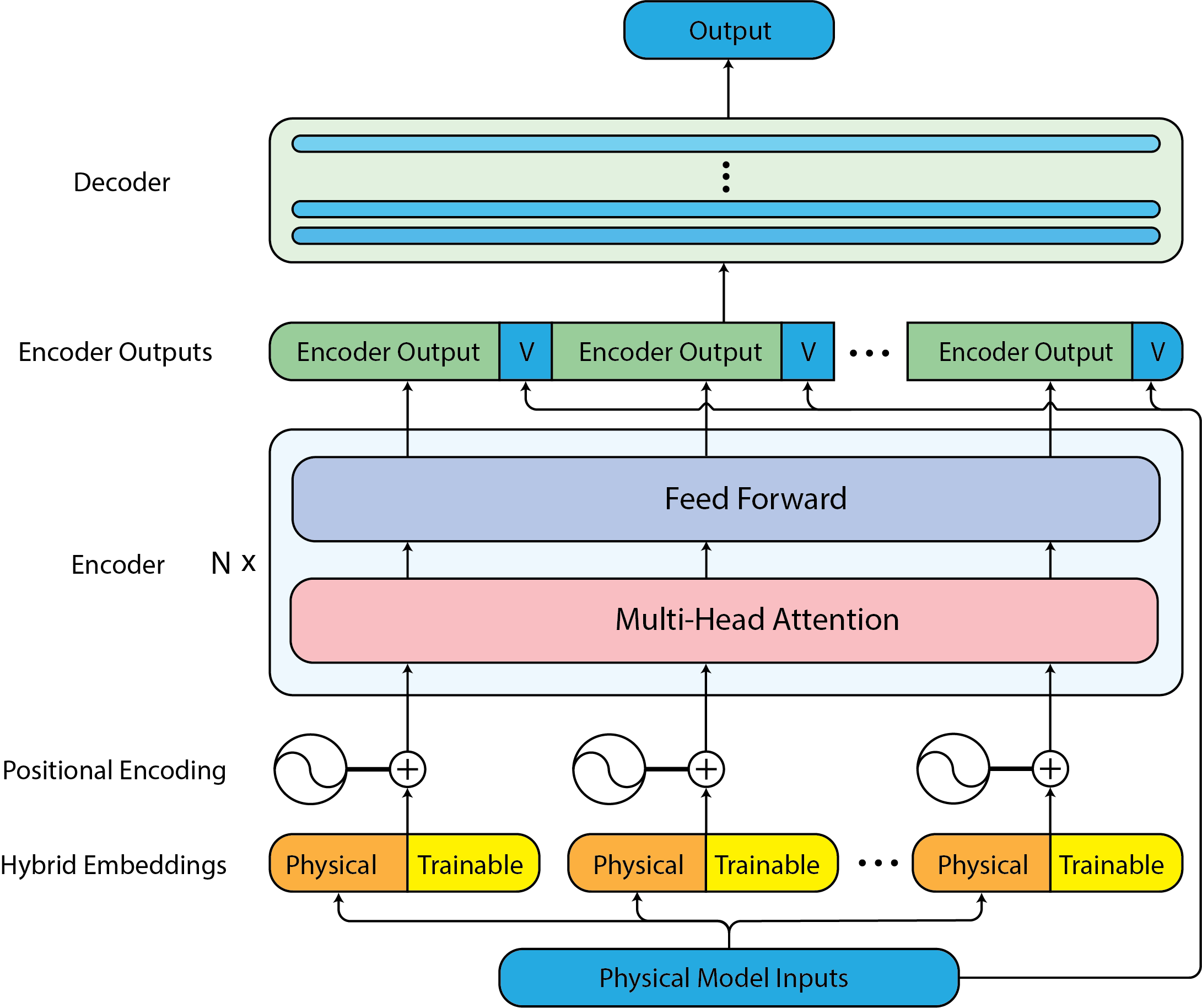}
    \caption{Transformer model for determining VCSEL parameters. The letter V denotes the initial normalized parameters of the model from inputs.}
    \label{fig:transformer_model}
\end{figure}

The architecture of our model is presented in Figure \ref{fig:transformer_model}. This is a modified transformer model adapted to solve the problem of determining the VCSEL radiation parameters based on the model parameters.

The input data consist of parameters from the physical model of the VCSEL, such as geometry and materials. 

Each physical quantity associated with the VCSEL is represented by what we call a "physical embedding." These embeddings contain vectors with information about specific physical quantities like energy, thickness, and refractive index. The inclusion of physical embeddings allows the model to retain critical quantitative details, which are pivotal for accurate simulations.

To extend the model's capability beyond static physical representations, trainable embeddings are also incorporated. This allows the model to adapt and identify patterns autonomously, creating what we refer to as hybrid embeddings. These are combinations of physical values and contextual information, enriching the model’s learning substrate and enabling a more dynamic adaptation to complex physical phenomena.

After the integration of hybrid embeddings, positional encoding is added to the mix. This step is essential for maintaining the sequence order of input data, especially crucial for transformer models where the relative positions of elements significantly influence the processing outcomes.

The processed data then enters the encoder, which consists of several layers, each with blocks of Multi-Head Attention and Feed Forward layers. The encoder extracts significant features from the data, with the Multi-Head Attention allowing the model to focus concurrently on different segments of the input sequence, thereby enhancing the detail and quality of information extraction. The Feed Forward layers are crucial as they enable the model to learn complex dependencies within the input data.

After encoding, the data are recombined with their original corresponding physical quantities. This recombination is vital for enhancing the model’s accuracy and stability as it progresses through the network, maintaining essential information about the physical parameters. All vectors are subsequently amalgamated into a single comprehensive vector that feeds into the decoder.

The decoder, a conventional sequential model consisting of several fully connected layers, processes the combined data. It predicts the final characteristics of the VCSEL, such as eigenenergy, quality factor, and threshold material gain. 

This architecture effectively models and predicts the physical parameters of complex structures, achieving high accuracy and adaptability in VCSEL characteristic prediction. This approach is a testament to our commitment to leveraging cutting-edge AI techniques to solve complex challenges in photonics engineering.

\section{Results and Discussion}
\label{sec:results_discussion}

Three configurations were developed for the transformer model. The first was engineered to calculate the two closest eigenvalues to a specified target energy, using the input energy to define the search region. The second configuration aimed to calculate the quality factor (Q-factor) using the energy eigenvalue, relevant to where the Q-factor had to be evaluated. The third configuration was used to determine the threshold material gain (TMG) $g_{th}$, employing the first eigenvalue. This TMG setup allowed the neural network to compute both the TMG and the corresponding energy, crucial since TMG is directly linked to the specific energy level at which the lasing threshold occurs. To accommodate different types of nanostructures, such as Fabry-Pérot, DBR, and VCSEL, the model is capable of handling various sequence lengths with the use of a masking technique in the transformer. This approach allows the model to adapt dynamically to the structural complexities of each type.

To ensure reliability and prevent overfitting, techniques such as dropout and k-fold cross-validation were implemented during training. Mean squared error (MSE) was utilized as the error metric.

\subsection{Eigenenergy Predictions}
\begin{figure}[htbp]
    \centering
    \includegraphics[width=\textwidth]{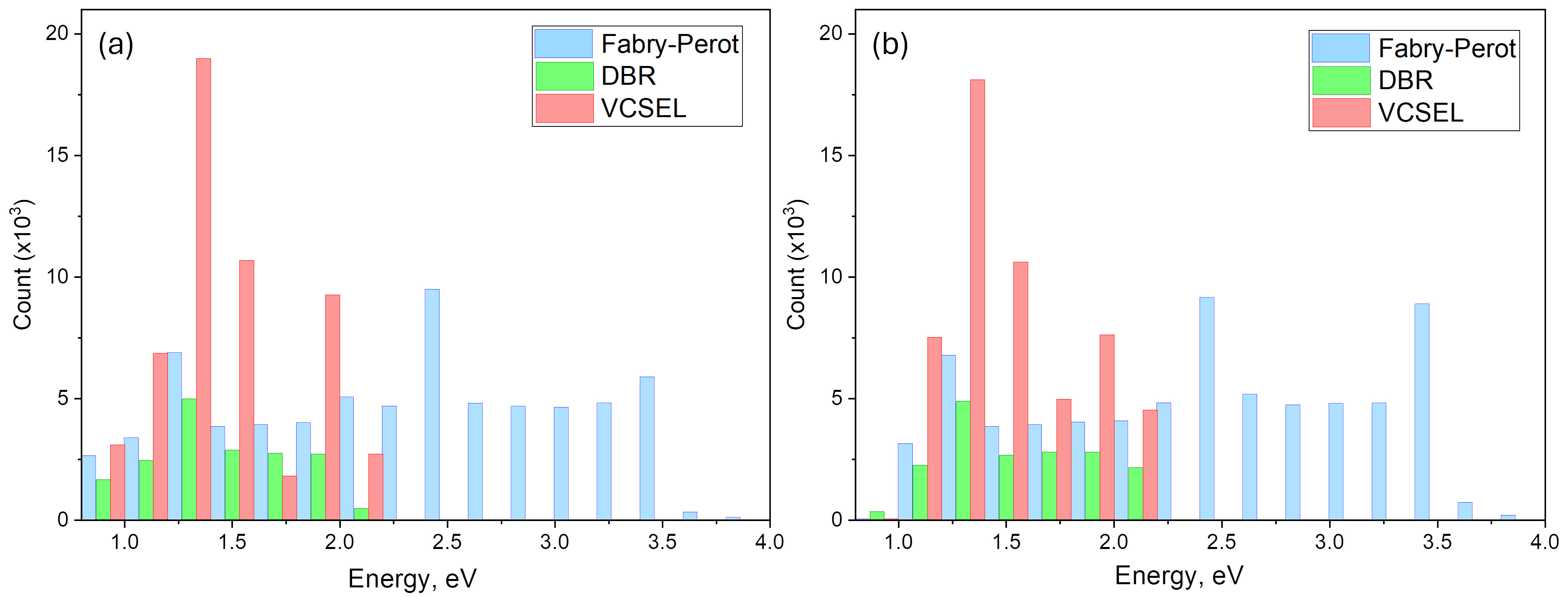}
    \caption{Histogram of eigenenergy distribution for three data types: single layers in blue, DBRs in green, and VCSELs in red.}
    \label{fig:eigenenergy_distribution}
\end{figure}

The distribution of the first two eigenenergies calculated for all data types in COMSOL is presented in Figure \ref{fig:eigenenergy_distribution}. The first and second identified eigenenergies are illustrated in Figures \ref{fig:eigenenergy_distribution}(a) and \ref{fig:eigenenergy_distribution}(b), respectively.

For DBRs and VCSELs, the eigenenergies generally did not significantly exceed $2$ eV, remaining closely within the initial energy ranges used for searching. The similarity in eigenenergy distributions for both modes can be attributed to the 2D axial symmetry considered in the simulations. In such symmetrical systems, there tend to be modes with similar eigenenergy but different azimuthal mode numbers.

\begin{figure}[htbp]
    \centering
    \includegraphics[width=\textwidth]{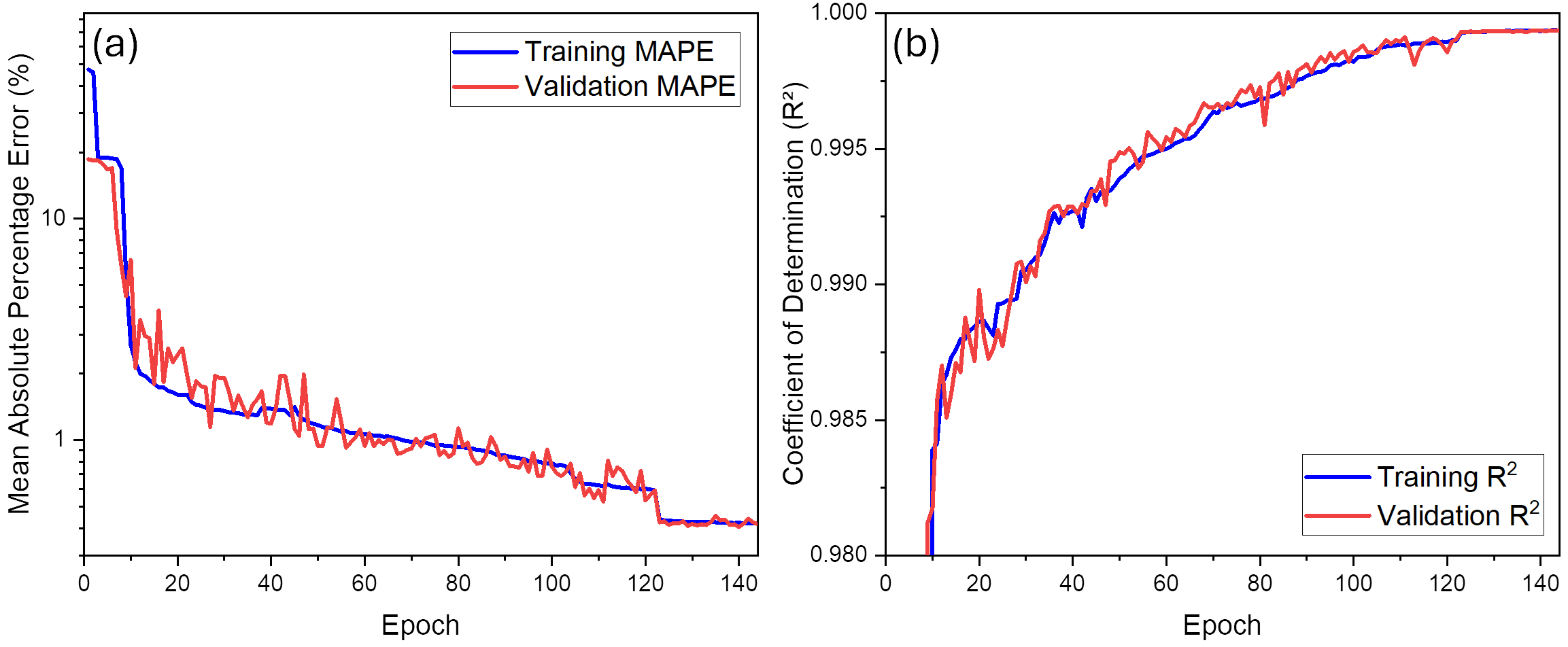}
    \caption{Training and Validation Metrics for Eigenenergy Prediction. (a) Mean Absolute Percentage Error (MAPE) with decreasing trends over training epochs for both training (blue) and validation (red) datasets. (b) Coefficient of determination (\( R^2 \)) indicating increasing model accuracy over epochs for both training and validation data.}
    \label{fig:training_validation_metrics}
\end{figure}

The results of training to predict the first two eigenenergies are presented in Figure \ref{fig:training_validation_metrics}. For this model, the transformer outputs two eigenenergies directly. Errors were calculated as the average error across both values. The graph in Figure \ref{fig:training_validation_metrics}(a) shows the mean absolute percentage error (MAPE) dependency on the epoch. By the $144$th epoch, the error reaches $0.4\%$. Figure \ref{fig:training_validation_metrics}(b) illustrates the dependence of the coefficient of determination $(R^2)$ on the epoch. It is also evident that by the $144$th epoch, the coefficient of determination reaches $0.9993$, which is very close to $1$, indicating that the model explains $99.93\%$ of the data variation, demonstrating high accuracy and reliability of the predictions. The learning rate also varied during the training process. 

For all data (comprising training, validation, and test data), the errors are as follows: Average MAPE Loss is $0.4\%$, and the standard deviation of the error $\sigma$ is $0.5\%$.

\begin{figure}[htbp]
    \centering
    \includegraphics[width=\textwidth]{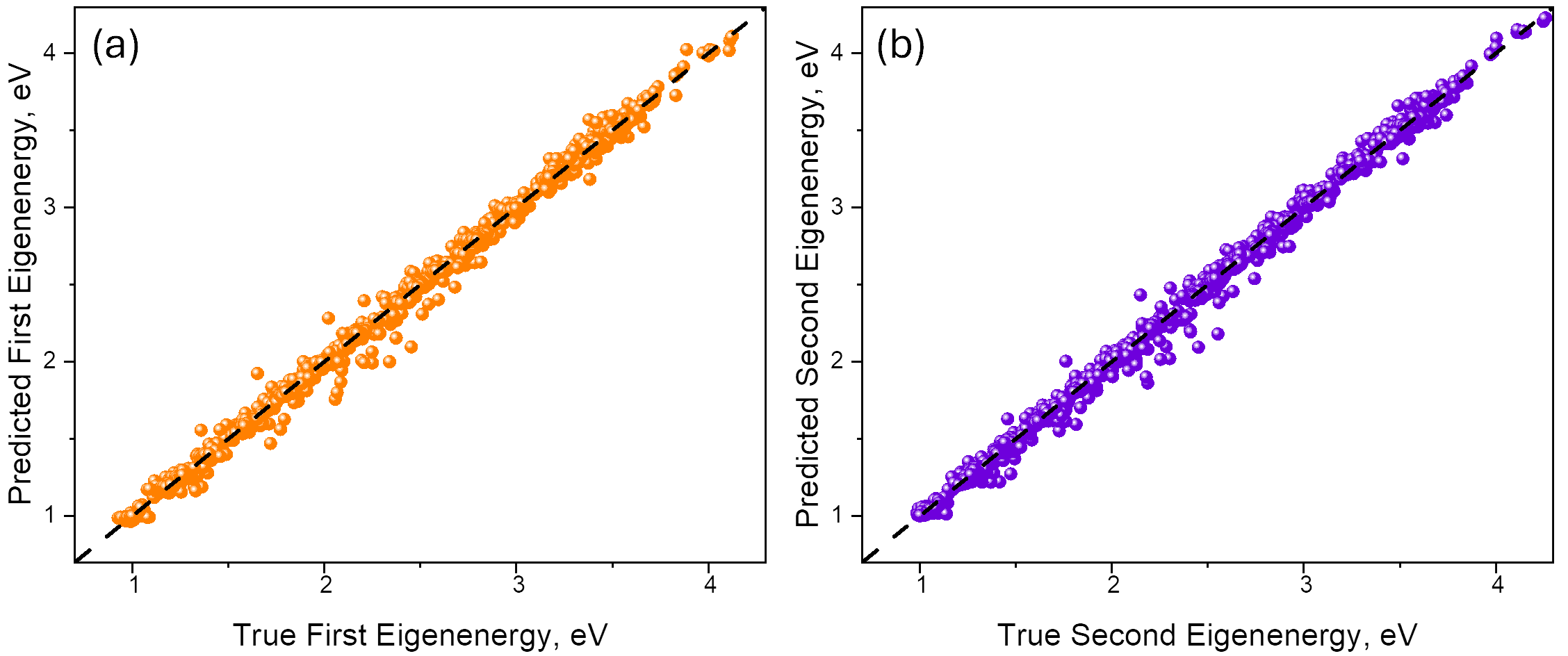}
    \caption{Comparison of the eigenenergies as computed in COMSOL Multiphysics and predicted by a Transformer model. The datasets consist of 14,031 test data points. (a) Distribution of the first eigenenergy. (b) Distribution of the second eigenenergy.}
    \label{fig:eigen_energy_comparison}
\end{figure}

Figure \ref{fig:eigen_energy_comparison} shows the distribution of true and predicted values for data that were not involved in the training process (test data). The figures depict (a) the distribution of the first eigenenergy and (b) the distribution of the second eigenenergy. It can be observed that across the entire training interval, the model accurately predicts the energy values for various VCSEL configurations.

The predictive performance of the transformer model in accurately determining the first two eigenenergies, along with its rapid convergence within just $144$ epochs, underscores the robustness and reliability of our approach. The minimal variation observed in the predicted data further reinforces the model's efficiency, making it a formidable tool in the study and optimization of VCSEL configurations.

\subsection{Quality factor predictions}
\label{subsec:quality_factor_predictions}

\begin{figure}[ht]
\centering
\includegraphics[width=\textwidth]{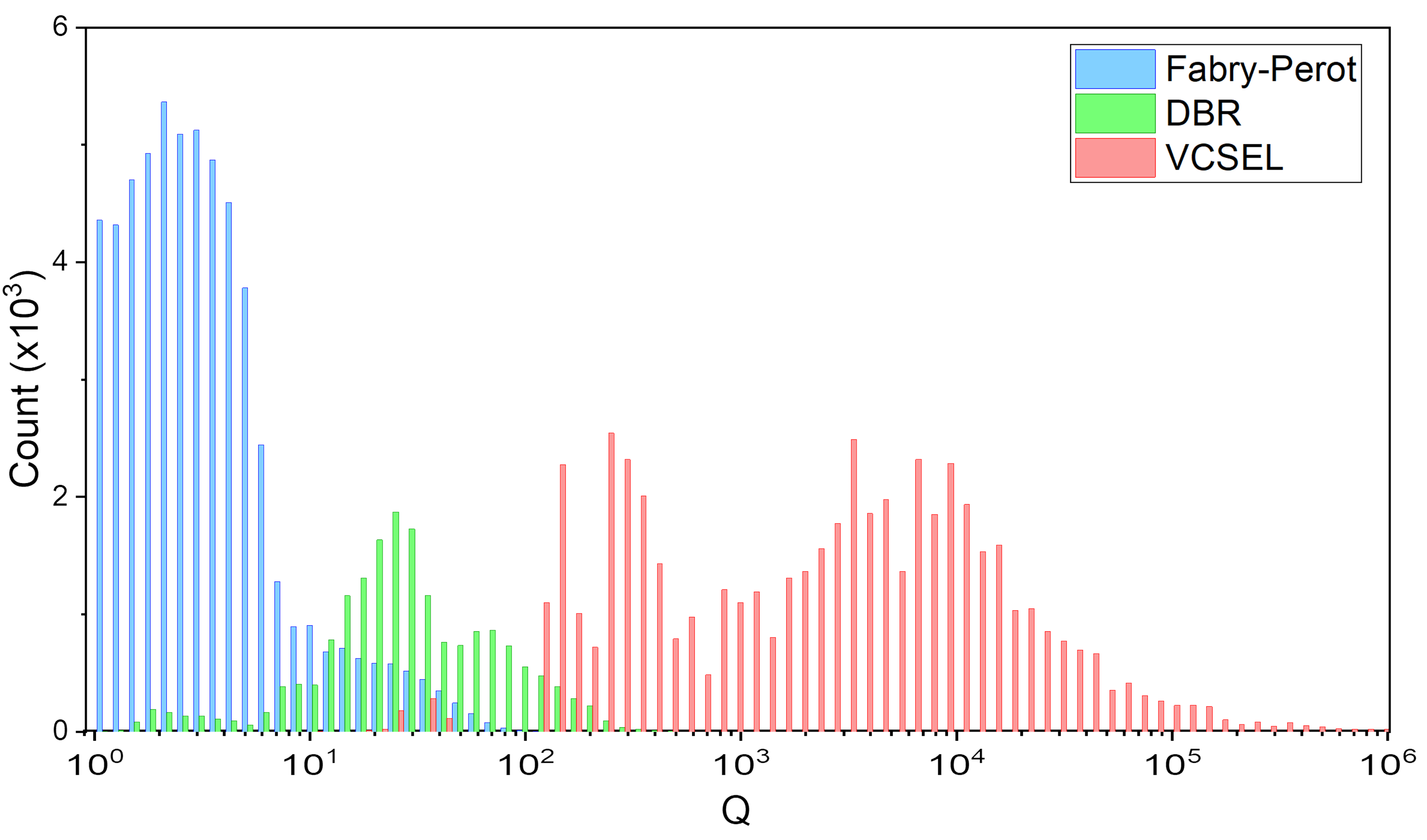}
\caption{Histogram of Q-factor distribution for three data types: single layers in blue, DBRs in green, and VCSELs in red.}
\label{fig:qfactor_distribution}
\end{figure}

Figure~\ref{fig:qfactor_distribution} displays the distribution of the logarithm of the Q-factor for all data types calculated in COMSOL, using a logarithmic scale for clarity.

The histogram shows that each data type exhibits distinct Q-factor ranges. Single layers typically have low Q-factors not exceeding $100$, DBRs range from $10$ to $500$, and VCSELs vary from $100$ to $10^6$, aligning with expectations for these structures. 

Following our successful eigenenergy predictions, we utilized our transformer model to forecast the quality factor (Q-factor) of VCSELs based on specified eigenenergy values and VCSEL configurations.

\begin{figure}[ht]
\centering
\includegraphics[width=\textwidth]{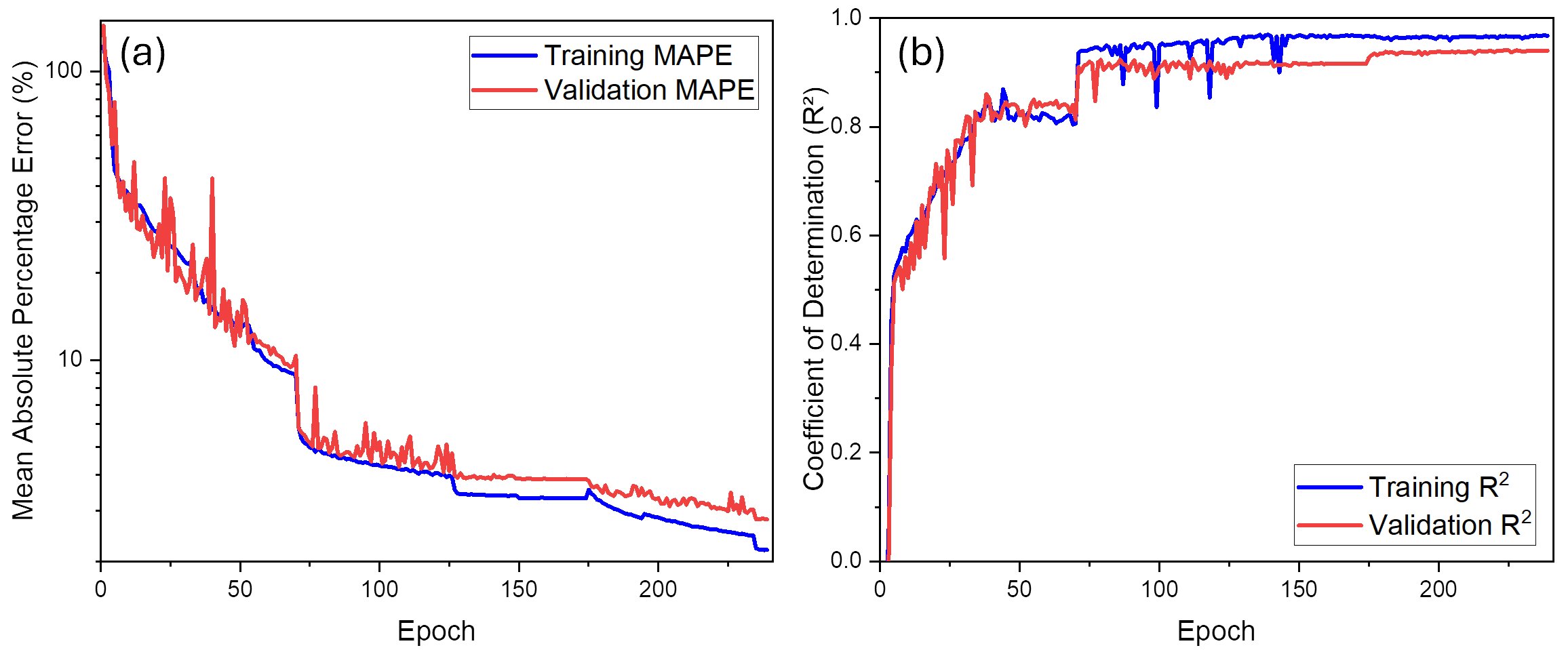}
\caption{Training and Validation Metrics for Q Prediction. (a) displays the Mean Absolute Percentage Error (MAPE) with decreasing trends over training epochs for both training (blue) and validation (red) datasets. (b) shows the coefficient of determination $(R^2)$, indicating increasing model accuracy over epochs for both training and validation data.}
\label{fig:q_prediction_training}
\end{figure}

The training outcomes for predicting the Q-factors are illustrated in Figure ~\ref{fig:q_prediction_training}(a), which displays the mean absolute percentage error over epochs. By the $240$th epoch, the error rates were $2.2\%$ for training data and $2.8\%$ for validation data. Figure ~\ref{fig:q_prediction_training}(b) shows the evolution of the coefficient of determination $(R^2)$, reaching $0.97$ for training data and $0.94$ for test data by the $240$th epoch.

Overall, for all data, the average MAPE Loss was $2.3\%$, with $\sigma$ at $4.8\%$.

\begin{figure}[ht]
\centering
\includegraphics[width=0.6\textwidth]{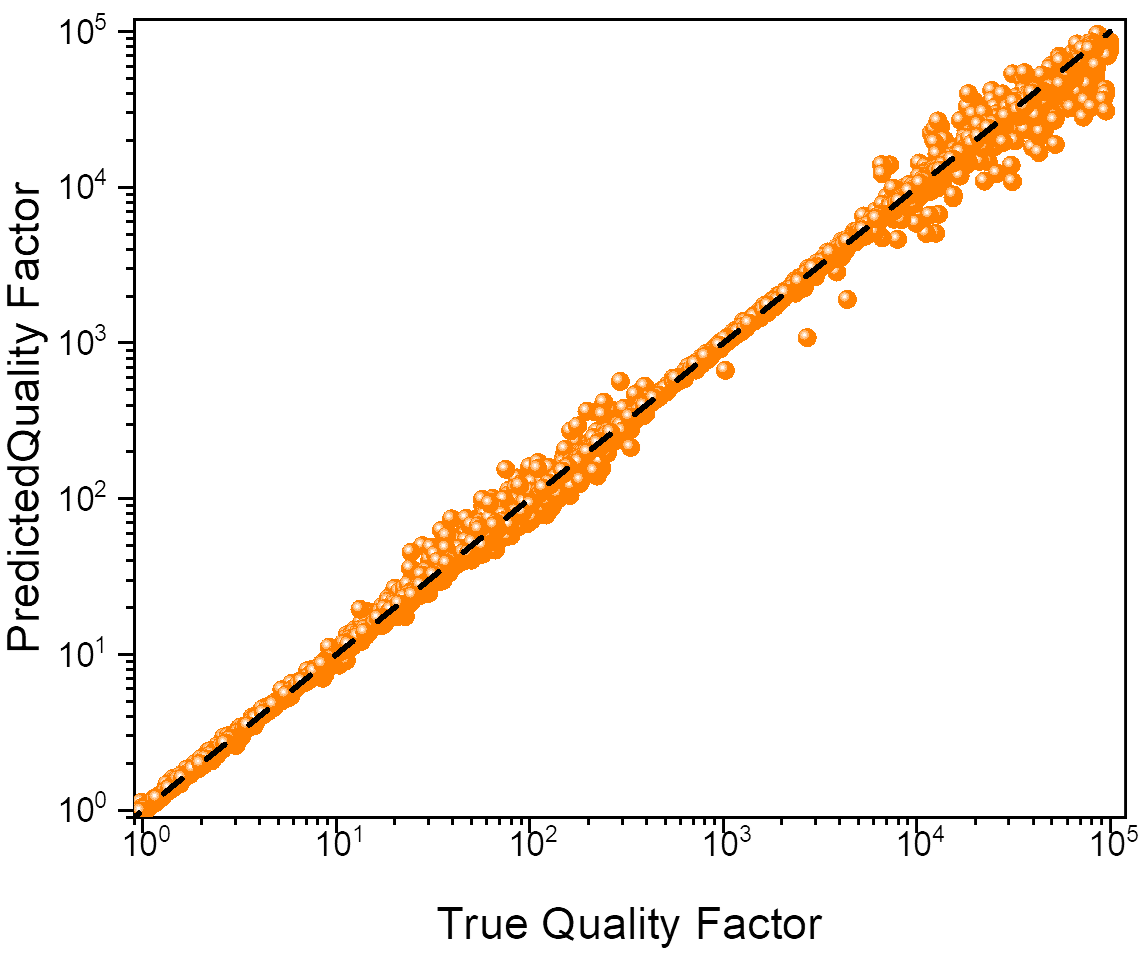}
\caption{Comparison of the quality factor as computed in COMSOL Multiphysics and predicted by a Transformer model. The datasets consist of 14,000 test data points.}
\label{fig:qfactor_predict}
\end{figure}

Figure ~\ref{fig:qfactor_predict} presents the distribution of true and predicted values for the test data, involving $14,000$ configurations. While the model required a slightly longer training period to achieve precision in Q-factor predictions, it efficiently computed results within a reasonable timeframe. Considering the complexity involved in predicting the Q-factor, which ranges broadly from $1$ to $10^5$, the model's performance is noteworthy.

This achievement underscores the transformer model's capability to handle complex, high-dynamic-range predictions effectively, further validating its utility in advanced VCSEL design and analysis, as well as other applications.

\subsection{Threshold material gain predictions}
\label{subsec:threshold_material_gain_predictions}

For the training, the model seeks both the energy and TMG since the generation energy does not always match the system's inherent energy. In the COMSOL model, finding the TMG involves an iterative process of searching both TMG and energy. For VCSELs, this iterative process results in energies that coincide with the eigenenergies due to the resonant structure of the devices. In contrast, for single layers, the iterative process may lead to optimal generation energies that do not align with the eigenenergies, lacking the resonant effect seen in multilayer structures. Thus, the model searches for both TMG and energy simultaneously.

\begin{figure}[ht]
\centering
\includegraphics[width=\textwidth]{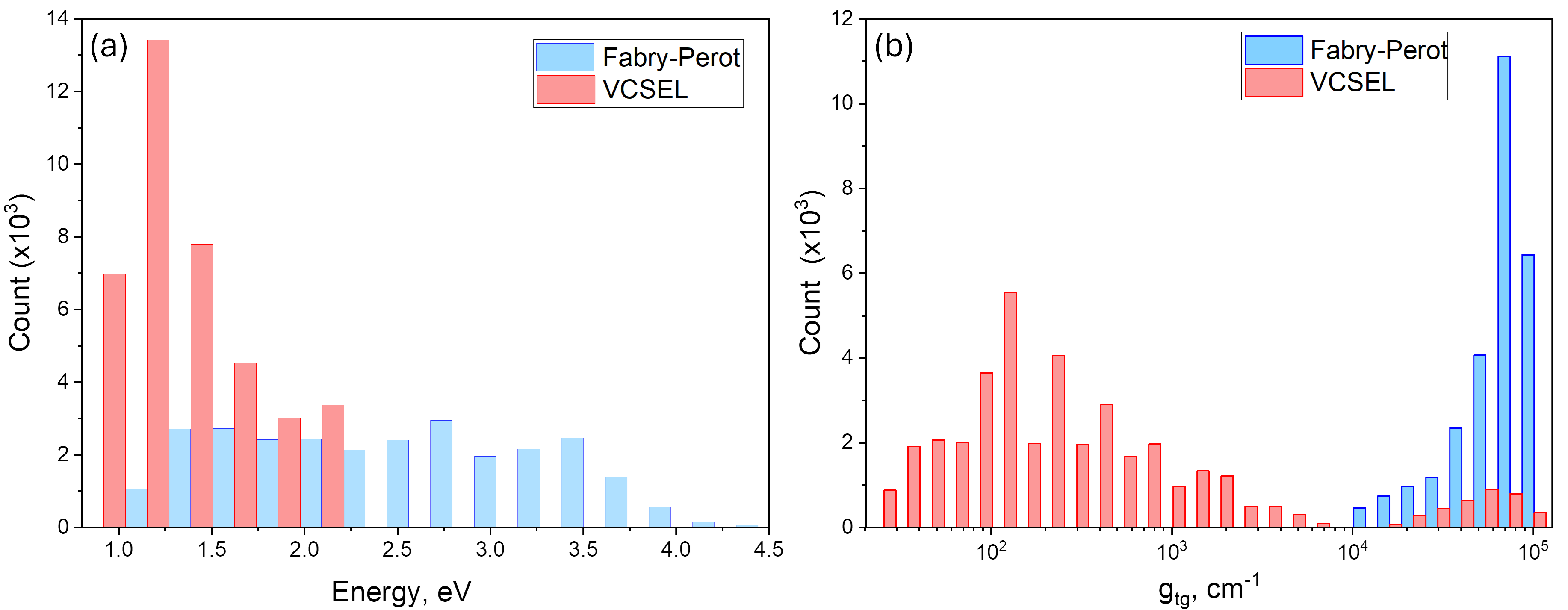}
\caption{Histogram of resonant energies (a) and threshold material gain (b) for two data types: fabri-perot in blue, and VCSELs in red.}
\label{fig:tmg_distribution}
\end{figure}

Histogram of resonant energies (a) and threshold material gain $g_{th}$ (b) is presented in Figure~\ref{fig:tmg_distribution}.

The data reveal that resonant energies for VCSEL systems are generally not much higher than $2 \, \text{eV}$. For threshold material gains, there is a clear separation between single layers and VCSELs, with single layers exhibiting a wide range from $10^4 \, \text{cm}^{-1}$ to $10^5 \, \text{cm}^{-1}$ and VCSELs typically below $1000 \, \text{cm}^{-1}$, although a small proportion of models show higher values up to $10^5 \, \text{cm}^{-1}$.

\begin{figure}[ht]
\centering
\includegraphics[width=\textwidth]{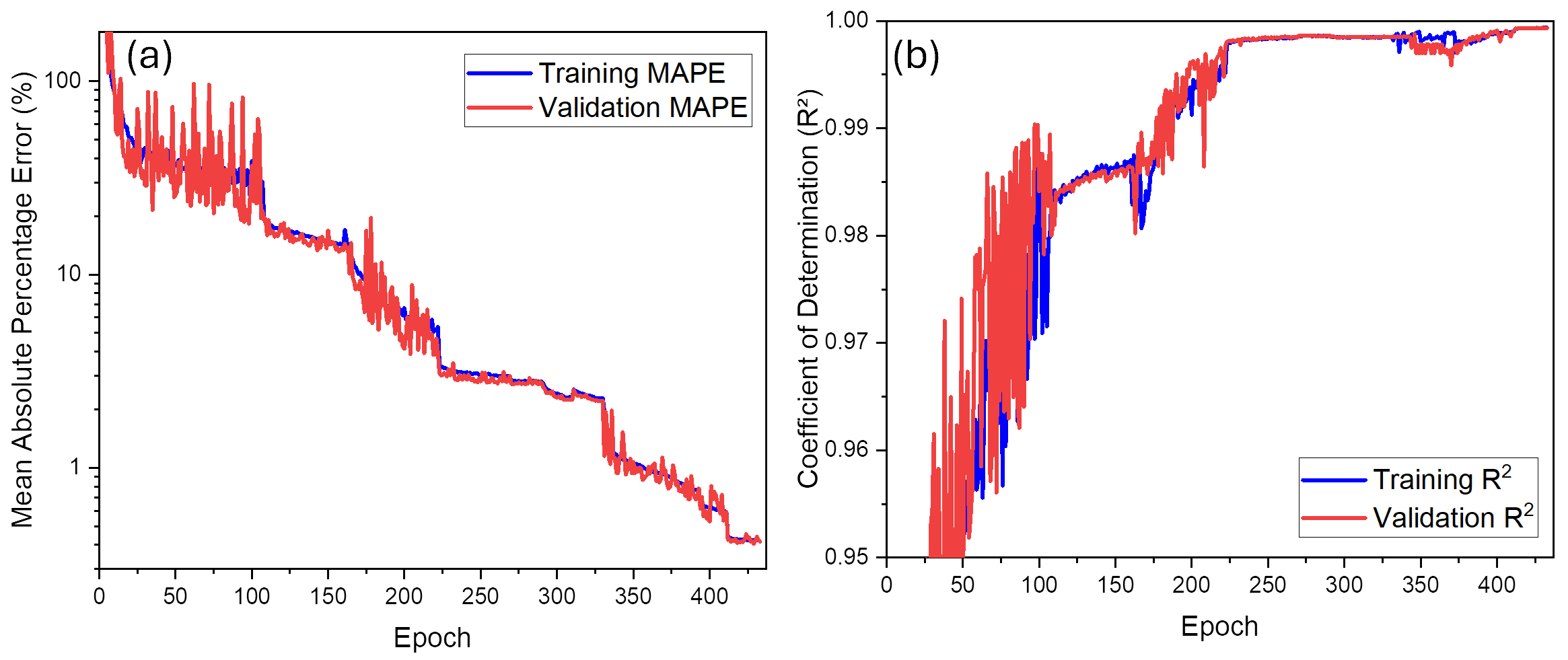}
\caption{Training and Validation Metrics for $g_{th}$ Prediction. (a) displays the Mean Absolute Percentage Error (MAPE) with decreasing trends over training epochs for both training (blue) and validation (red) datasets. (b) shows the coefficient of determination \( R^2 \), indicating increasing model accuracy over epochs for both training and validation data.}
\label{fig:tmg_train}
\end{figure}

The training results for finding $g_{th}$ and resonant energies are presented in Figure ~\ref{fig:tmg_train}(a). It shows the dependence of the mean absolute percentage on the epoch. By the $433$rd epoch, the error reaches $0.4\%$ on average for both metrics. Figure ~\ref{fig:tmg_train}(b) displays the relationship of the coefficient of determination $(R^2)$ over the epochs. It is also evident that by the $433$rd epoch, the coefficient of determination reaches $0.9993$, which is very close to $1$, indicating that the model explains $99.93\%$ of the data variation.

For all data, the errors are as follows: the average Mean Absolute Percentage Error (MAPE) is $0.6\%$ for energies and $1.0\%$ for $g_{th}$, while the standard deviation of the error $\sigma$ is $1.2\%$ for energy and $1.4\%$ for $g_{th}$. However, if we calculate the errors only for complete VCSEL models, excluding single-layer models, the errors are as follows: the average MAPE is $0.12\%$ for energies and $0.66\%$ for $g_{th}$, and the standard deviation of the error $\sigma$ is $0.09\%$ for energy and $0.3\%$ for $g_{th}$.

\begin{figure}[ht]
\centering
\includegraphics[width=\textwidth]{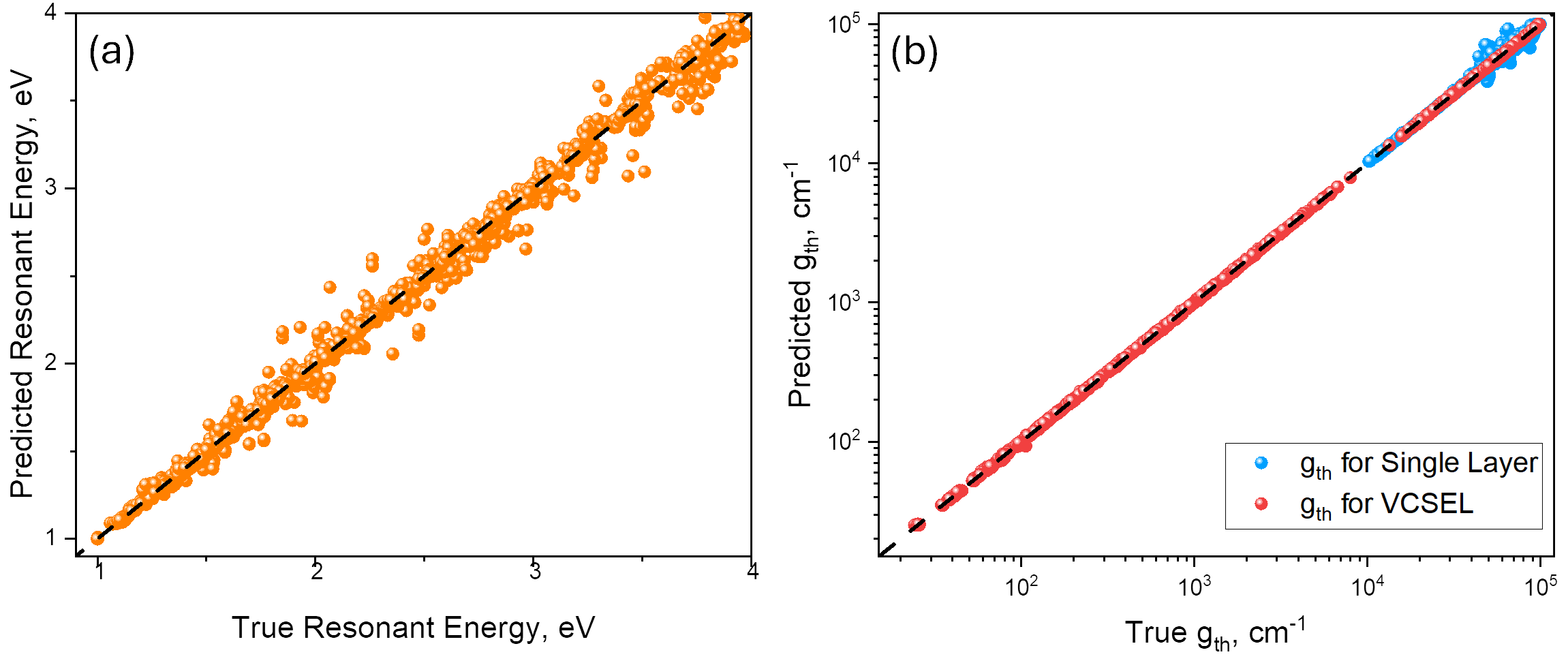}
\caption{Comparison of the resonant energies and \( g_{\text{th}} \) as computed in COMSOL Multiphysics and predicted by a Transformer model. The datasets consist of 6,650 test data points. (a) Distribution of the resonant energy. (b) Distribution of the threshold material gain.}
\label{fig:tmg_predict}
\end{figure}

Figure ~\ref{fig:tmg_predict} presents the distribution of true and predicted values for the test data, involving approximately six and a half thousand configurations. 

In Figure~\ref{fig:tmg_predict}(a), the distribution of resonant energies demonstrates that the predicted values closely match the actual data. In Figure~\ref{fig:tmg_predict}(b), the prediction errors for TMG are shown for both single-layer models and complete VCSEL models. It is evident that the errors for the VCSEL models are significantly lower, and the data points are much more tightly aligned with the axis of true values.

During this study, a performance comparison was conducted between the proposed transformer model and a traditional multilayer neural network, which was tasked with predicting quality factors. Despite significant efforts to optimize the training of the multilayer neural network, the minimum error achieved did not fall below $230\%$. This outcome highlights the complexity of the task and the potential advantages of using the transformer architecture for such types of challenges, especially considering its ability to efficiently handle data with high variability and complex structure.

The successful application of the transformer model to predict VCSEL emission properties has not only demonstrated high accuracy but also underscored the potential of deep learning in photonics. Through various configurations and experiments, the transformer model has proven to be robust, capable of handling complex and high-dimensional data effectively, and providing reliable predictions that are critical for optimizing VCSEL designs.

To further enhance prediction accuracy and its application in the industry, one could consider using conformal predictions or a two-level model. In the framework of a two-level model, the first part would predict a range of values, and the second, trained within this range, would refine the results. This approach can provide additional accuracy and reliability of predictions, which is especially important for industrial applications.

\section{Conclusions}
\label{sec:conclusions}
In this study, we have introduced and validated a novel approach using transformer-based neural networks to predict VCSEL emission properties, marking a significant advancement over traditional simulation methods. This approach is faster, more accurate, and computationally efficient, making it a game-changer for VCSEL technology.

The transformer model achieved exceptional accuracy in predicting key parameters such as eigenenergy, quality factor, and threshold material gain. This precision is critical for the design and optimization of VCSELs, demonstrating the model's high accuracy and efficiency. By comparing the transformer model with traditional multilayer neural networks, we underscored its superior capability in handling complex photonics simulations. 

The success of this model in VCSEL technology also has broader implications for photonics. It opens the door to applying this innovative approach in other areas of photonics where similar challenges in simulation and design exist. 

In conclusion, the application of transformer-based neural networks represents a significant leap forward in the field of photonics, particularly in the modeling and simulation of VCSELs. Our work demonstrates the practical benefits of this technology and lays a solid foundation for its broader application in the field, setting the stage for future advancements and innovations.

\bibliographystyle{unsrt}  

\appendix
\renewcommand{\thetable}{A\arabic{table}}
\setcounter{table}{0}

\section*{\LARGE Appendix}
\section{Additional Embeddings}
\label{app:other_embeddings}

Here, we include tables similar to Table 1 for other structures such as VCSEL and single layer configurations.

\textbf{Single Layer Embeddings} 

Table \ref{tab:single_layer_embeddings} presents the embeddings for the single layer configurations.

\begin{table}[ht]
    \centering
    \caption{Physical embeddings for single layer.}
    \label{tab:single_layer_embeddings}
    \renewcommand{\arraystretch}{1.2} 
    \setlength{\arrayrulewidth}{0.3mm} 
    \setlength{\tabcolsep}{4pt}
    \newcolumntype{C}[1]{>{\centering\arraybackslash}m{#1}} 
    
    \begin{tabular}{C{0.5cm} | C{1cm} C{1.5cm} C{1cm} C{1cm} C{1cm} C{1.5cm} C{1cm} C{1cm} C{2cm}}
    
        & \textit{Energy} & \textit{Thickness} & \textit{N} & \textit{K} & \textit{GAIN} & \textit{Boundary} & \textit{DBR} & \textit{Pairs} & \textit{Value} \\ \hline
        \rowcolor{green!10} 1 & 1 & 0 & 0 & 0 & 0 & 0 & 0 & 0 & 1.2 \\
        2 & 0 & 0 & 1 & 0 & 0 & 1 & 0 & 0 & 1.0 \\
        \rowcolor{green!10} 3 & 0 & 0 & 0 & 1 & 0 & 1 & 0 & 0 & 0.0 \\
        4 & 0 & 1 & 0 & 0 & 0 & 0 & 0 &  0 & 90 \\
        \rowcolor{green!10} 5 & 0 & 0 & 1 & 0 & 0 & 0 & 0 &  0 & 1.6 \\
        6 & 0 & 0 & 0 & 0 & 1 & 0 & 0 &  0 & 5000 \\
        \rowcolor{green!10} 7 & 0 & 0 & 1 & 0 & 0 & 1 & 0 &  0 & 1.0 \\
        8 & 0 & 0 & 0 & 1 & 0 & 1 & 0 &  0 & 0.0 \\
    \end{tabular}
\end{table}

The first embedding describes an energy level of $1.2$ $eV$. The next two embeddings describe the lower medium with parameters $n$ and $k$. Following this, three embeddings describe the active layer: its $thickness$, $n$, and $gain$ values. Finally, the last two embeddings describe the properties of the upper semi-infinite medium, their $n$ and $k$ values.

\textbf{VCSEL Embeddings}

Table \ref{tab:vcsel_embeddings} presents the embeddings for the VCSEL configurations.

\begin{table}[ht]
    \centering
    \caption{Physical embeddings for VCSEL.}
    \label{tab:vcsel_embeddings}
    \renewcommand{\arraystretch}{1.2} 
    \setlength{\arrayrulewidth}{0.3mm} 
    \setlength{\tabcolsep}{4pt}
    \newcolumntype{C}[1]{>{\centering\arraybackslash}m{#1}} 
    
    \begin{tabular}{C{0.5cm} | C{1cm} C{1.5cm} C{1cm} C{1cm} C{1cm} C{1.5cm} C{1cm} C{1cm} C{2cm}}
    
        & \textit{Energy} & \textit{Thickness} & \textit{N} & \textit{K} & \textit{GAIN} & \textit{Boundary} & \textit{DBR} & \textit{Pairs} & \textit{Value} \\ \hline
      \rowcolor{green!10} 1 & 1 & 0 & 0 & 0 & 0 & 0 & 0 & 0 & 1.6 \\
        2 & 0 & 0 & 1 & 0 & 0 & 1 & 0 & 0 & 2.5 \\
        \rowcolor{green!10} 3 & 0 & 0 & 0 & 1 & 0 & 1 & 0 & 0 & 0.0 \\
        4 & 0 & 0 & 0 & 0 & 0 & 0 & 1 & 1 & 30 \\
        \rowcolor{green!10} 5 & 0 & 1 & 0 & 0 & 0 & 0 & 1 & 0 & 131.51 \\
        6 & 0 & 0 & 1 & 0 & 0 & 0 & 1 & 0 & 1.47 \\
        \rowcolor{green!10} 7 & 0 & 0 & 0 & 1 & 0 & 0 & 1 & 0 & 0.0 \\
        8 & 0 & 1 & 0 & 0 & 0 & 0 & 1 & 0 & 82.59 \\
        \rowcolor{green!10} 9 & 0 & 0 & 1 & 0 & 0 & 0 & 1 & 0 & 2.35 \\
        10 & 0 & 0 & 0 & 1 & 0 & 0 & 1 & 0 & 0.0 \\
        \rowcolor{green!10} 11 & 0 & 1 & 0 & 0 & 0 & 0 & 0 & 0 & 131.51 \\
        12 & 0 & 0 & 1 & 0 & 0 & 0 & 0 & 0 & 1.47 \\
        \rowcolor{green!10} 13 & 0 & 0 & 0 & 1 & 0 & 0 & 0 & 0 & 0.0 \\
        14 & 0 & 1 & 0 & 0 & 0 & 0 & 0 & 0 & 162.67 \\
        \rowcolor{green!10} 15 & 0 & 0 & 1 & 0 & 0 & 0 & 0 & 0 & 2.35 \\
        16 & 0 & 0 & 0 & 1 & 0 & 0 & 0 & 0 & 0.0 \\
        \rowcolor{green!10} 17 & 0 & 1 & 0 & 0 & 0 & 0 & 0 & 0 & 5.0 \\
        18 & 0 & 0 & 1 & 0 & 0 & 0 & 0 & 0 & 2.35 \\
        \rowcolor{green!10} 19 & 0 & 0 & 0 & 0 & 1 & 0 & 0 & 0 & 1200 \\
        20 & 0 & 1 & 0 & 0 & 0 & 0 & 0 & 0 & 162.67 \\
        \rowcolor{green!10} 21 & 0 & 0 & 1 & 0 & 0 & 0 & 0 & 0 & 2.35 \\
        22 & 0 & 0 & 0 & 1 & 0 & 0 & 0 & 0 & 0.0 \\
        \rowcolor{green!10} 23 & 0 & 0 & 0 & 0 & 0 & 0 & 1 & 1 & 25 \\
        24 & 0 & 1 & 0 & 0 & 0 & 0 & 1 & 0 & 131.51 \\
        \rowcolor{green!10} 25 & 0 & 0 & 1 & 0 & 0 & 0 & 1 & 0 & 1.47 \\
        26 & 0 & 0 & 0 & 1 & 0 & 0 & 1 & 0 & 0.0 \\
        \rowcolor{green!10} 27 & 0 & 1 & 0 & 0 & 0 & 0 & 1 & 0 & 82.59 \\
        28 & 0 & 0 & 1 & 0 & 0 & 0 & 1 & 0 & 2.35 \\
        \rowcolor{green!10} 29 & 0 & 0 & 0 & 1 & 0 & 0 & 1 & 0 & 0.0 \\
        30 & 0 & 0 & 1 & 0 & 0 & 1 & 0 & 0 & 1.0 \\
        \rowcolor{green!10} 31 & 0 & 0 & 0 & 1 & 0 & 1 & 0 & 0 & 0.0 \\
    \end{tabular}
\end{table}

The first embedding describes an energy. Two embeddings $2$ and $3$ describe the lower semi-infinite medium. From embeddings $4$ through $10$, the lower Distributed Bragg Reflector (DBR) is detailed. Embeddings $11$ through $13$ add another layer for symmetry. Embeddings $14$ to $16$ and $20$ to $22$ describe the lower and upper parts of the cavity, respectively. The active region, or quantum well, is covered by embeddings $17$ to $19$. Embeddings $23$ to $29$ detail the upper DBR. Finally, embeddings $30$ and $31$ describe the properties of the upper semi-infinite layer.

This arrangement ensures a comprehensive representation of the VCSEL structure, providing detailed information about each component and its position within the overall device architecture.

\end{document}